# Fuzzy logic based approaches for gene regulatory network inference

**Khalid Raza**

*Department of Computer Science, Jamia Millia Islamia, New Delhi, India.*
Email: kraza@jmi.ac.in

April 26, 2018

**Abstract:** The rapid advancement in high-throughput techniques has fuelled the generation of large volume of biological data rapidly with low cost. Some of these techniques are microarray and next generation sequencing which provides genome level insight of living cells. As a result, the size of most of the biological databases, such as NCBI-GEO, NCBI-SRA, is exponentially growing. These biological data are analyzed using computational techniques for knowledge discovery – which is one of the objectives of bioinformatics research. Gene regulatory network (GRN) is a gene-gene interaction network which plays pivotal role in understanding gene regulation process and disease studies. From the last couple of decades, the researchers are interested in developing computational algorithms for GRN inference (GRNI) using high-throughput experimental data. Several computational approaches have been applied for inferring GRN from gene expression data including statistical techniques (correlation coefficient), information theory (mutual information), regression based approaches, probabilistic approaches (Bayesian networks, naïve byes), artificial neural networks, and fuzzy logic. The fuzzy logic, along with its hybridization with other intelligent approach, is well studied in GRNI due to its several advantages. In this paper, we present a consolidated review on fuzzy logic and its hybrid approaches for GRNI developed during last two decades.

## 1. Introduction

The field of Bioinformatics is one of the youngest and growing among the modern sciences. Computational systems biology is a sub-discipline of bioinformatics which deals with the dynamic studies of interactions of biological macromolecules. From the last few decades, a wide variety of methods and concepts borrowed from mathematics, computer science, statistics and probability theory and applied in the area of bioinformatics and computational systems biology. Among these methods, the fuzzy logic theory also has lots of potential applications in different areas of bioinformatics, including Microarray gene expression analysis, gene biomarkers, and gene regulatory network inference (GRNI). The fuzzy logic theory is needed to solve several challenging problems in bioinformatics which are beyond the capabilities of other existing approaches. Today, fuzzy logic has major applications in the area of engineering and technologies. However, it has relatively minor applications in the area of bioinformatics and biomedical sciences. According to PubMed biomedical literature repository, total number of publications with "fuzzy logic" in title or abstract during 1964-2017 is 4,609 (Fig. 1). However, total number of publications with "fuzzy" in title or abstract during the same period is 9,250. Last decade has witnessed an average of more than 500 papers per year (Fig. 1). In the coming years, it is expected to rapidly grow visibility and applications of fuzzy-logic-based techniques for bioinformatics and medical research due to its several promises (Xu et al. 2008).

The GRNI from high-throughput gene expression data is a well-posed challenge from last few decades. Several computational methods have been proposed ranging from simple statistical approaches, such as correlation, mutual information (Margolin et al., 2006; Raza & Parveen, 2013b), to sophisticated methods such as Bayesian network, Petri net, artificial neural networks (ANN), fuzzy logic to name a few. DREAM challenge (http://dreamchallenges.org) was posed to develop accurate GRNI methods for small and mid size networks, but inference of large GRN is still a challenge. Few reviews and tutorials exist in the area such as modeling, simulation and analysis of GRN (De Jong, 2002; Schlitt & Brazma, 2007; Cho et al., 2007; Karlebach & Shamir, 2008; Lee & Tzou, 2009; Yaghoobi et al., 2012; Chai et al., 2014; Al Qazlan et al., 2015), soft computing approaches (Fogelberg & Palade, 2009; Mitra et al., 2011; Raza & Parveen, 2013a), evolutionary approaches (Sîrbu et al., 2010; Raza & Parveen, 2012), data integration approaches (Hecker et al., 2009; Chen & VanBuren, 2012; Wani & Raza, 2018) and comparative genomics approaches (Thompson et al., 2015). The purpose of this review paper is to systematically present fuzzy logic based approaches to model and inference of GRN developed in the last two decades. The paper is organized as follows. Section 2 presents an introduction to GRN. Section 3 presents a brief over of fuzzy logic and generic steps of a fuzzy system. Section 4 and its subsections review fuzzy logic and its hybridization with other computational techniques for GRNI. Finally, section 5 present discussion and conclusion.

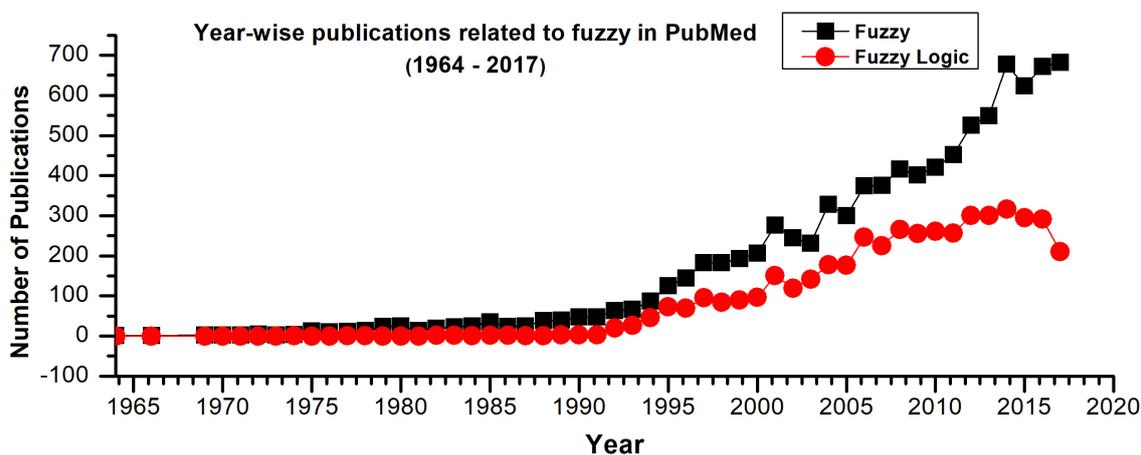

**Fig. 1** Year-wise publications related to fuzzy in PubMed biomedical repository

## 2. Gene Regulatory Networks

All living bodies are made up of cells which in-house DNA inside its nucleus. Genes are made up of DNA, the information house of the cell, which carries genetic blueprint used to make proteins. Every single gene stores a particular set of instruction which codes for a specific protein. The process of conversion of genes into a functional product (protein) is known *gene expression*. Gene expression process is tightly regulated which lets a cell to respond to its changing environment. Genes act as switching circuit – ON or OFF. When a gene is turned off, it no longer provides the directions for making proteins. Gene expression is considered as the most fundamental level where genotype gives rise to the phenotype. Cell regulates the expression of genes in response to changes occurred in the environment which give rise to regulatory network (Raza, 2014; Raza, 2016a).

A gene regulatory network (GRN) consists of set of genes interacting to each other to control a specific cell function. GRNs play pivotal role in development, differentiation and response to environment (Raza, 2016b). A GRN is like a directed graph consisting of nodes and edges, where nodes represent genes and their regulators, and edges represent their regulatory relationships such as activation or inhibition (Raza & Parveen, 2013a; Raza & Jaiswal, 2013). The regulatory genome resemble as a logic processing system, which receives multiple inputs and processes them as combinations of logical functions such as "AND", "OR", or "switch" functions (Davidson & Levine, 2005). Identification of GRNs within a cell is very important because it has direct influence on the development and survival of living organisms. Recent advancement in high-throughput techniques such as microarray and next-generation sequencing (NGS) have generated huge amount of Transcriptomic data (Raza & Alam, 2016). Last few decades has witnessed the development of plenty of computational methods for the inference of GRN from gene expression profiles, also known as reverse-engineering or reconstruction of GRN. A general introduction to GRN and their applications in clinical and personalized medicine can be found in Filkov (2005) and Emmert-Streib et al. (2014), respectively.

## 3. Fuzzy Logic: A Brief Overview

The concept of Fuzzy Logic (FL) was introduced by Lotfi Zadeh in 1965 with the introduction of fuzzy set theory (Zadeh, 1965). It gave birth to a new mechanism to process data by using partial set membership rather than crisp set membership. Till late 1970s, fuzzy set theory was not applied to control systems because of limited processing power of computers. Zadeh reasoned that we do not need precise, numerical inputs, and yet it is capable of highly adaptive control. Fuzzy logic is a problem solving control system methodology which allows to include vagueness, uncertainty, imprecision and partial truth in computing problems, and provides an effective solution for conflict resolution of multiple criteria. Fuzzy logic tends itself to system ranging from simple, small embedded system to large, complex problem such as control system, knowledge-base system, image processing, power engineering, robotics, industrial automation, consumer electronics, multi-objective optimization, weather forecasting, stock trading, medical diagnosis and treatment, bioinformatics and so on (Singh et al., 2013). It can be implemented both at hardware and software level.

Fuzzy logic uses a simple rule-based "IF X AND Y THEN Z" approach for solving control problem. It offers several unique features making it a good choice for several modeling and control applications: i) it is inherently robust to imprecise and noisy inputs, ii) it may be programmed to fail safely, iii) it can process any reasonable number of inputs and generate numerous outputs, and iv) capable of modeling nonlinear systems easily which may be sometime difficult or impossible to model mathematically. Fuzzy logic based modeling and control system consists of three major steps: fuzzification, inference rule, and defuzzification (Fig. 2).

## *Fuzzification*

Fuzzification provides a way to transform precise quantitative values (e.g. temperature=40°C, 25°C or 10°C) into qualitative (nominal) values (e.g. temperature="High", "Medium" or "Low"). There are other ways to transform precise values into discrete descriptors, but FL offers a systematic and unbiased way without the need of expert knowledge about the system. The complete range of input data is first measured and then it is fuzzified into discrete subsections using some appropriate *membership functions* (MFs). The MF represents the magnitude of participation and associates a "weight" with each input, defines functional overlaps between inputs. The membership function is used to map the non-fuzzy inputs to fuzzy linguistic terms and vice-versa. Sometimes, before fuzzification, a normalization technique is applied to scale all the numeric values in the given range which helps to prevent attributes with large ranges. Triangular, trapezoidal and Gaussian shape are the most common MFs, but there are other forms of the MFs such as singleton, Piece-wise linear and exponential (Mendel, 1995; Woolf & Wang, 2000).

## *Fuzzy Inference*

In this step, a rule base in the form of "IF-THEN" rule is constructed in order to control the output variable. The purpose of inference engine is to draw conclusions from rule base. Fuzzy set operations (AND, OR, or NOT) are applied to evaluate fuzzy rules and combine their results. After evaluating result of each rule, individual results are combined (using accumulation methods such as *maximum*, *bounded sum* or *normalized sum*) to obtain a final output.

## *Defuzzification*

The inference step gives the results as fuzzy values that need to defuzzified to get a final crisp output. The *defuzzifier* component performs the defuzzification according to the MF of the output variable. Some of the commonly used methods for deffuzzification are *center of sums (COS)*, *center of gravity (COG)*, *weighted average*, and *maxima method*, and so on (Mendel, 1995).

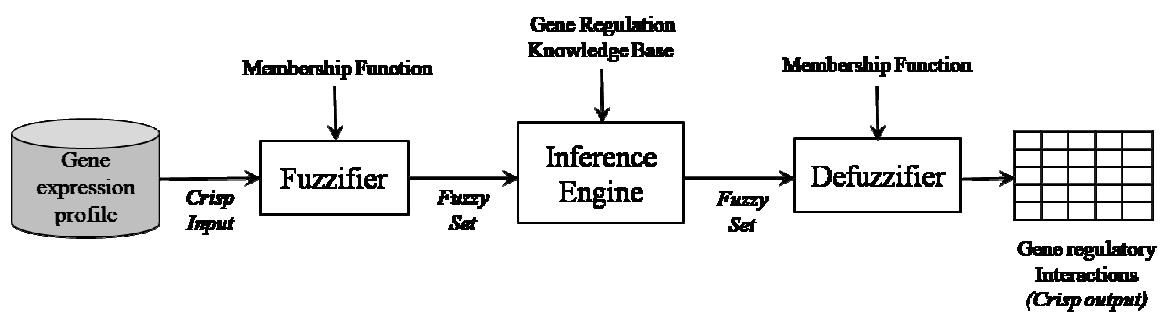

**Fig. 2** A generic pipeline of fuzzy logic model of GRN inference

# 4. Fuzzy Logic based GRN inference methods

The biological systems are very complex which behave in a fuzzy manner. Fuzzy logic offers a mathematical framework for modeling, describing and analyzing biological systems (Raza, 2016a). Following subsections describe different fuzzy based approach for GRN inference.

## *4.1 Classical Fuzzy Logic Model*

Fuzzy logic, being capable to represent nonlinear systems and incorporate domain knowledge using fuzzy rules, has been used for modeling GRNs and gene expression analysis. Woolf & Wang (2000) mentioned three main advantages of fuzzy logic in gene expression studies, (i) it extracts trends rather than precise values, and therefore it can inherently handle noises in the gene expression data; (ii) its predicted results are easily interpretable because decision rules are casted in the form of "if-then" rules like the language used in day-to-day conversation; and (iii) it is computationally efficient and scalable to virtually unlimited number of components.

### *Woolf and Wang's Algorithm*

One of the initial successful attempts to apply fuzzy logic for GRNI was done by Woolf and Wang (Woolf & Wang, 2000), who proposed a novel algorithm to find gene triplets in the form of activators (A), repressors (R), and targets (T) in yeast. The capability of fuzzy logic to deal with the subjectivity of quantitative measurements using membership functions it can define quantitative set of rules to model GRN. This algorithm assumes three states of gene expression: LOW, MEDIUM, and HIGH. For instance, in gene expression analysis one such rule may be "for a gene target T regulated by an activator A and repressor R, if A's expression level is LOW and R's expression is HIGH, this will imply that T's expression will be LOW". Similarly, if A is HIGH and R is LOW then T will be HIGH (Fig. 3). The membership functions and set of rules defined by Woolf and Wang is shown in Fig. 4.

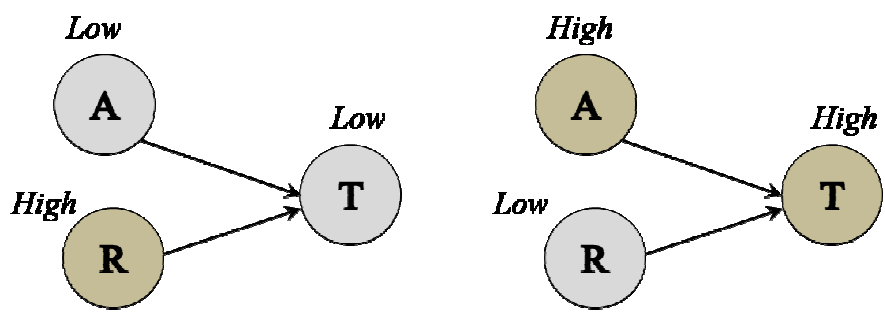

**Fig. 3** General example of rules for Activator (A)-Repressor (R)-Target (T) relations

The fuzzy logic algorithm for GRN inference has following steps:
*Step 1 Fuzzification:* The gene expression data is converted to fuzzy values by first scaling it between 0 and 1, and then normalized value is converted into different membership classes (such as LOW, MEDIUM, or HIGH) using membership function shown in Fig. 4(A). For instance, if normalized expression value is 0.25 then fuzzified value would be 0.5 LOW, 0.5 MEDIUM, and 0 HIGH. Similarly, normalized expression value of 0.5 and 0.75 will have fuzzified membership values as (LOW=0, MEDIUM=1, HIGH=0) and (LOW=0, MEDIUM=0.5, HIGH=0.5), respectively.

*Step 2 Creation and Comparison of Triplets:* After fuzzification process, all possible gene triplets (A-R-T) are defined as expression values of corresponding three genes and compared using decision matrix mentioned in Fig. 4(B). Fuzzified values of A and R are entered into the decision matrix and, at points where their prediction overlap, a score is generated as fuzzified value of predicted T. Hence, predicted expression values of T for all the time-point is computed.

*Step 3 Defuzzification:* The predicted fuzzy values of T for all the time-points are then defuzzified into a crisp value.

*Step 4 Triplet Screening:* The predicted value of T for all the time-points is compared with that of the observed T expression values. For each triplet, $r^2$ between the predicted T and the observed T is calculated to check concurrence with assertion in the rule table. Triplets having a low $r^2$ fit the assertion better and are considered as highly confident triplets. Triplets with $r^2 < 0.015$ are screened which correspond to an average error of 3% or less. Although, this algorithm was computationally expensive which took ~10 days to find A-R-T triplets among 1,898 genes, but it leads to several further improvements and extensions.

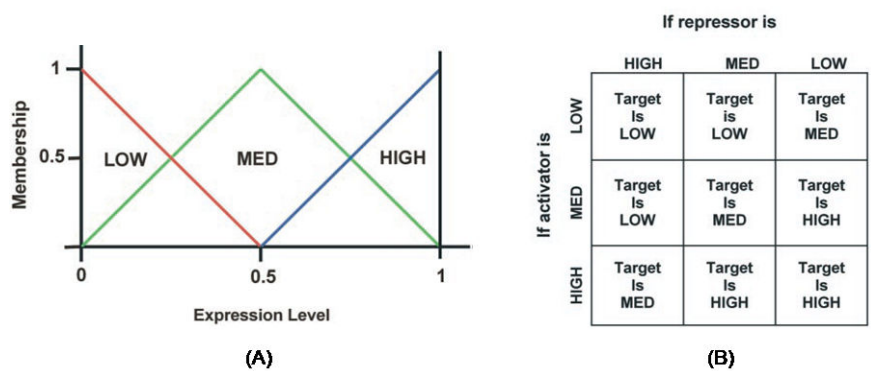

**Fig. 4 (A)** Membership functions and **(B)** decision rule matrix used by Woolf & Wang (2000)

In 2003, Ressom and collaborators (Ressom et al., 2003a; Ressom et al., 2003b) improved the performance of Woolf and Wang algorithm by reducing computing time up to 50% and accommodating co-activators and co-repressors in the GRN model. Reduction in computation time was achieved by introducing clustering as a preprocessing step which reduces the number of gene combinations to be analyzed without any effect on the results. In 2006, Ram et al. (2006) also extended Woolf and Wang's model by considering two important assumptions: (i) input transcript factors (TFs) are driver for gene expression, and therefore inputs having lower gene expressions are assumed to produce no significant change at output gene expression level; (ii) similar gene expression profiles are redundant for computation, and therefore these are grouped to reduce computation cost. The method attempts to eliminate false positives from the classical fuzzy model proposed by Woolf & Wang (2000), and also reduce the redundant computation and make the algorithm faster. Ma &

Chan (2008) proposed fuzzy logic based mining techniques to define fuzzy dependency among genes and discover hidden fuzzy dependency relationships in high-dimensional time-series gene expression data.

In a complex network, relatively dense regions are termed as *network modules* which represent a set of regulated genes corresponding to similar biological function. Mahanta et al. (2014) developed a fuzzy network module extraction technique (FUMET). The FUMET takes two input parameters such as number of modules and membership threshold, and works on weighted co-expression network. Based on user input, FUMET infers biologically important and highly co-expressed modules.

Mostly, existing fuzzy logic based approaches obtain a qualitative model of the systems, and are unable to cope with the quantitative response of the system. Also, unavailability of kinetic data is a major barrier in the quantitative modeling. Bordon et al. (2015) presented a fuzzy logic based approach which quantitatively model the behaviour of a biological system even through kinetic data are uncertain or partially known. They performed the demonstration of their proposed method on a three-gene repressilator model.

Beside development of fuzzy-based models of GRNI, there are several application of fuzzy logic based framework for discovering novel GRN and pathways and other applications. Dickerson et al. (2001) applied fuzzy cognitive maps to model metabolic networks. Nodes of the map represent biomolecules such as genes, proteins, RNAs, and other small molecules, or various stimuli, and edges represents regulatory and metabolic relationships. Bosl (2007) applied fuzzy rule-based method representing expert knowledge in cell cycle regulation and tumor growth. They examined several common network motifs and constructed fuzzy rule-based model of hedgehog regulation of cell cycle. Aldridge et al. (2009) applied fuzzy logic framework to study the kinase pathway crosstalk in TNF, EGF, and insulin receptors of colon carcinoma cells in human. They also uncovered several other relationships between genes, such as MK2 and ERK pathways, unexpected inhibition of IKK following EGF treatment. Brock et al. (2009b) stated that fuzzy logic and related techniques can be applied as a screening tool for GRN detection. Jin et al. (2009) empirically investigated influence of regulation logic on the dynamics of GRN motifs consisting of three genes having positive and negative feedback loops. Fuzzy logic framework is also applied to model Lambda switch – a widely studied paradigm of gene regulation (Laschov et al., 2009).

### *4.2 Fuzzy Cognitive Maps*

Fuzzy cognitive maps (FCMs), first introduced by Bart Andrew Kosko (Kosko, 1986), is one of the most powerful intelligent tools, considered as soft computing techniques, which combines features from fuzzy logic and ANN. It works well for modeling complex processes in different field of studies (Amirkhani et al., 2017).

Fuzzy cognitive Maps (FCMs) are diagraphs consisting of nodes and weighted edges that can model causal flow between biomolecules such as genes and proteins in a GRN. It has the capability to cope with lack of quantitative information as how various biomolecules interact. The genes or proteins are represented as causal fuzzy sets and degree to which they are dependent on each other. A FCM consists of $N$ nodes and weighted edges $w_{ij} \in [-1, 1]$ between them, where each node represents a concept to be modeled. An example of FCM with 3 nodes is shown in Fig. 5. In FCM, the values of the nodes can be represented as a vector,

$$C = \{C_1, C_2, ..., C_N\} \tag{1}$$

where $C_i \in [0, 1]$ is fuzzy membership that works as activation degree for gene *i* in the GRN. The activation degree $C_i(t)$ of gene *i* at time point *t* is computed as,

$$C_i(t) = f\left(\sum_{j=1}^{N} w_{ji} C_i(t-1)\right) \tag{2}$$

where *f*(.) is a transfer function which is usually a sigmoidal function.

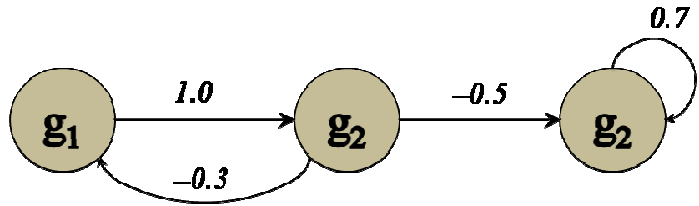

**Fig. 5** An example of GRN as FCM consisting of three genes as nodes and weighted edges as their casual regulatory relation

Du et al. (2005) attempt to model gene-gene interaction (also known as interaction parameter) within a GRN as fuzzy functions using FCMs, where interactions stand for causal flow. They first clustered data by multi-scale fuzzy k-means algorithm and then search for weighted time correlation between the cluster centre time profiles. Their method consists of three steps:

*(i) Multiscale Fuzzy K-Means Clustering:* This step clusters the gene expression data at different level based on expression similarity.

*(ii) Construction of GRN:* After clustering similar gene expression profiles (i.e., co-regulated genes), it finds the relationships among co-regulated genes. If two genes have similar expression profiles, they may have three possible relationships, (a) they are co-regulated by other genes, (b) first gene regulates the second or vice versa, or (c) there is no causal relationship between them. If $x_A$ and $\tau_A$ be the gene expression and regulation time delay of gene A at time t, respectively; and $w_{BA}$=[0, 1] be weight representing inference of gene B to A, and $b_A$ be the bias of gene A (i.e. default gene expression without regulation), the GRN model can be represented as a simplified linear model (D'Haeseleer et a., 1999; Du et al., 2005),

$$x_A(t + \tau_A) = \sum_B w_{BA} x_B + b_A \tag{3}$$

The discrete time correlation between genes A and B ($R_{AB}$) can be represented as (Du et al., 2005),

$$R_{AB}(\tau) = \sum_n x'_A(n) x'_B(n-\tau) \tag{4}$$

where $x'_A$ and $x'_B$ are the standardized expression profiles for genes A and B, and $\tau$ is the time shift. The combined time correlation for multiple datasets can be computed as (Du et al., 2005),

$$R^C_{AB}(\tau) = \sum_k w_k R^k_{AB}(\tau) \tag{5}$$

where $w_k$ and $R^k_{AB}(\tau)$ are weight and time correlation result of $k^{th}$ dataset, respectively. Given a time correlation threshold $\theta$, there is significant regulation between genes or clusters if,

$$max|R^C_{AB}(\tau)| > \theta \tag{6}$$

Assuming clusters as node and significant links as edges of the GRN, Du et al. (2005) defined four types of regulations:

a) Positive regulation between A and B, if $R^C_{AB}(\tau') > 0$, $\tau' \neq 0$
b) Negative regulation between A and B, if $R^C_{AB}(\tau') < 0$, $\tau' \neq 0$
c) A and B are positively co-regulated, if $R^C_{AB}(\tau') > 0$, $\tau' = 0$
d) A and B are negatively co-regulated, if $R^C_{AB}(\tau') < 0$, $\tau' = 0$

where $\tau'$ is time delay between expression profiles between gene A and B. The sign of $\tau'$ indicates the direction of regulation. For example, $\tau' > 0$ means gene B regulates gene A, and $\tau' < 0$ means gene A regulates gene B.

*(iii) Network Evaluation:* The last step is the evaluation of inferred network using a fuzzy metrics. The validity and strength of the interaction is evaluated using evidence strength and co-occurrence of similar gene function within a cluster. Each gene within a cluster is weighted using Gaussian window function. Gene Ontology (GO) annotation database is used to calculate the fuzzy measures based on gene functions within a cluster. Based on strength of supporting evidence, the GO terms of each cluster are weighted.

Traditionally, construction of FCMs relies on domain knowledge. However, several attempts have been made to automatically learn FCMs and discover the domain knowledge in the form of causal relation between genes from the data. An attempt to apply FCM with Ant Colony Optimization (ACO) for GRN model was carried out by Chen et al. (2012). The relations between genes are modeled as fuzzy relations in FCMs, which avoids discretization of gene expression data. Chen et al. (2012) proposed ACO based learning algorithm to learn FCMs, where the optimization problem is decomposed into several small problems that makes the algorithm scalable. The algorithm was tested on DREAM-4 project (Stolovitzky et al., 2009) datasets of 10 genes and 100 genes networks which shows promising results. FCMs based approaches for automatic learning of FCMs from data suffers from several limitations: (i) the learned FCMs are in small-scale, (ii) its accuracy is relatively very low, (iii) high computational complexity, and (iv) density of learned FCMs is very high (Wu & Liu, 2017). Few attempts have been made to accurately and robustly learn large-scale FCMs from small amount of data without any prior knowledge (Chen et al., 2015; Wu & Liu, 2016; Liu et al., 2017; Zou & Liu, 2017; Wu & Liu, 2017). Chen et al. (2015) applied decomposed genetic algorithm to learn large-scale FCMs based GRN, which is found to perform better than other decomposition framework such as ACO, differential evolution, and PSO, even for small and noisy datasets. To robustly train FCMs from noisy data, convex optimization methods such as least absolute shrinkage and selection operator (LASSO), called LASSO-FCM (Wu & Liu, 2016) and compressed sensing (CS) (Donoho, 2006), called CS-FCM (Wu & Liu, 2017) are applied successfully. Both methods, LASSO-FCM and CS-FCM, decompose the FCMs learning problem into sparse signal reconstruction problems which is solved by LASSO and CS, respectively, and tested on synthetic data (DREAM3 and DREAM4) of varying sizes and density. CS-FCM performs well for network having 1,000 or more nodes, and also needs less data for its construction. Further, other decomposition-based models to train large-scale FCMs with decomposition (FCMD), such as dynamical multi-agent genetic algorithm (dMAGA), called dMAGA-FCMD (Liu et al., 2016; Liu et al., 2017), and mutual information (MI) based two-phase memetic algorithm (MA), called MIMA-FCM (Zou & Liu, 2017) are also developed which claim to perform well on large-scale FCMs learning. Also, for FCM-based reconstruction of GRN, memetic algorithm (MA) combined with ANN, called MANN-FCM-GRN is proposed by Chi & Liu (2016), where MA is used to find regulatory interactions, and ANN is applied to estimate GRN interaction strength.

*4.3 Dynamic Fuzzy Models*

Dynamic fuzzy modeling approach has the capability to incorporate the prior structural knowledge to the GRN model and infer gene interactions as fuzzy rules. Sun et al. (2010) applied this technique to model GRNs and extracted gene interactions as easily interpretable fuzzy rules. They used a T-S fuzzy model with *m* fuzzy rules as,

$$R^l: IF\ z_1(k)\ is\ M_1^l,\ z_2(k)\ is\ M_2^l, \ldots z_r(k)\ is\ M_r^l, THEN$$
$$y(k+1) = \hat{A}_l(q^{-1})y(k) + \hat{B}_l(q^{-1})u(k) + \hat{d}_l,\ l = 1,2,\ldots m \quad (7)$$

where,
$R^l = l^{\text{th}}$ fuzzy inference rule,
m = number of inference rules,
$M_r^l$ = fuzzy sets,
$u(k) \in R^g$ = input external stimuli variables which influence gene regulation
$y(k) \in R^p$ = output gene expression variables,
$z(k) \in R^r$ = premises measurement variables,

$(\hat{A}_l, \hat{B}_l, \hat{d}_l) = l^{th}$ local model with shift operator $q^{-1}$ defined by $q^{-1}y(k) = y(k-1)$

$$\hat{A}_l(q^{-1}) = \hat{A}_{l1} + \hat{A}_{l2}q^{-1} + \cdots + \hat{A}_{ln_y}q^{-n_y+1}$$

$$\hat{B}_l(q^{-1}) = \hat{B}_{l1} + \hat{B}_{l2}q^{-1} + \cdots + \hat{B}_{ln_u}q^{-n_u+1}$$

It is important to note that Sun et al. (2010) has lumped all the information into single matrix $\hat{A}_l(q^{-1})$ for simplicity. The local fuzzy model shown in equation (7) only represents properties of GRN in local region. Therefore, center-average defuzzifier and inference can be applied to equation (7) as,

$$y(k+1) = \hat{A}(q^{-1}, \mu(z))y(k) + \hat{B}(q^{-1}, \mu(z))u(k) + \hat{d}_l(\mu(z)) \qquad (8)$$

where,

$$\hat{A}(q^{-1}, \mu(z)) = \sum_{l=1}^{m} \mu_l \hat{A}_l(q^{-l}) \qquad (9)$$

$$\hat{B}(q^{-1}, \mu(z)) = \sum_{l=1}^{m} \mu_l \hat{B}_l(q^{-l}) \qquad (10)$$

$$\mu(z) = (\mu_1, \ldots, \mu_m) \qquad (11)$$

The dynamic fuzzy GRN model proposed by Sun et al. (2010) as shown in equation (8) can be applied to represent non-linear relationship among genes. In addition, a generalized fuzzy clustering approach is also applied to incorporate prior structural knowledge which helps in faster convergence of the model and find optimal number of fuzzy rules. There are two important characteristics of this model: (i) prior structural knowledge can be included into the model, and (ii) non-linear dynamic properties of regulatory network can be well captured. For detailed discussion, refer Sun et al. (2010).

### *4.4 Neuro-Fuzzy Hybrid*

The learning and adaptation feature of artificial neural network (ANN) can be combined with fuzzy logic to infer GRN. Neuro-fuzzy is one of the mostly applied hybrid approach used for GRNI. Jung & Cho (2007) applied neuro-fuzzy inference system consisting of two modules, a neuro-fuzzy inference module (NFIM) and an evolutionary strategy learning module (ESLM) (Fig. 6). The NFIM contains ANN whose link weights are assigned fuzzy rules. The ESLM trains and optimize the NFIM parameters. Firstly, gene expression profile is converted into a mapping form and then it is mapped into NFIM by training ESLM. Finally, fuzzy rules of NFIM infer the regulatory relations between genes. Gene expression profile mapped to fuzzy rules makes NFIM noise tolerant.

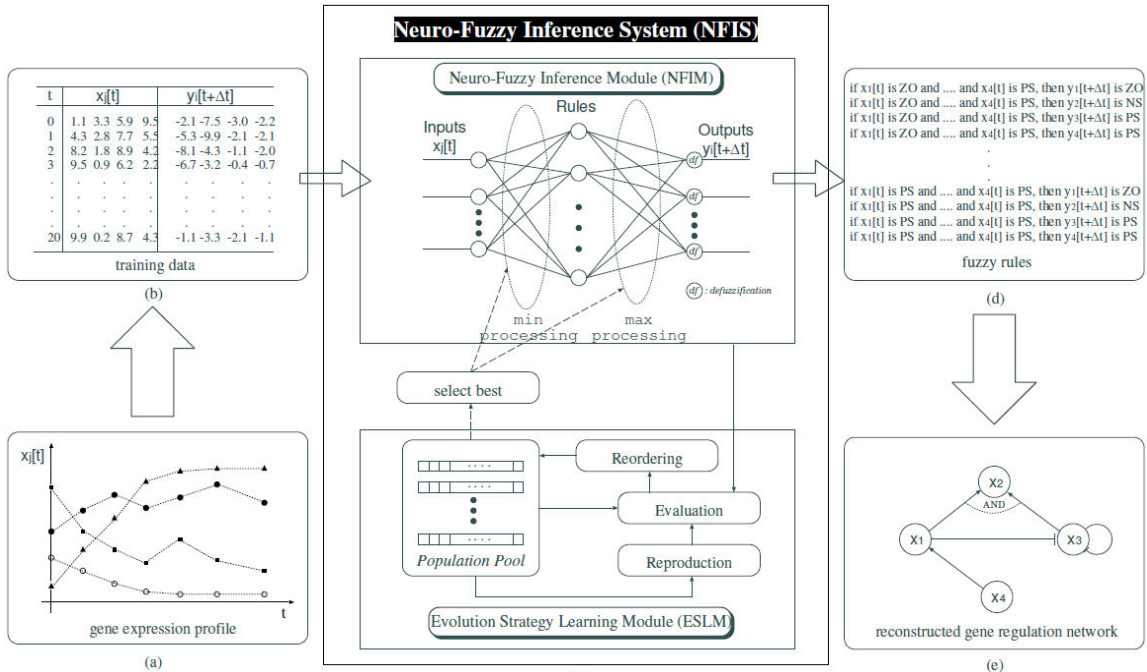

**Fig. 6** Neuro-Fuzzy Inference System consisting of two modules: NFIS and ESLM (Jung & Ch, 2007)

Maraziotis et al. (2007) presented a multilayer evolutionary trained neuro-fuzzy recurrent network (ENFRN) which infers the complex causal relationships between genes, and thus, determines potential regulators and regulation type of target genes as fuzzy rules. They adopt Zadeh-Mamdani's fuzzy model of fuzzy inference, and used six-layer RNN architecture. The dimension of the first layer (input layer) and the sixth layer (output layer) are equal to the number input variables and number of output variables, respectively. The fourth layer represents fuzzy rules and the number of nodes is equal to the number of fuzzy rules. The dimensions of rest of the layers are selected automatically using structure learning algorithm of ENFRN. Munoz et al. (2009) combined the features of ordinary differential equation (ODE) based models to fuzzy inference system (FIS), called ODE-FIS, and trained it through ANN. The GRN adapt the membership and output function from FIS. Datta et al. (2009) attempt to apply combination fuzzy membership and RNN (Fuzzy RNN), and determined the interaction parameters between the genes. The method treats the weights between neurons as the gene-gene interactions parameter values. The connection weights are represented using fuzzy membership and differential evolution algorithm is applied to determine optimal membership distribution of weights. Further, the membership distribution is defuzzified by using centroidal defuzzification method.

Liu et al. (2011) applied neuro-fuzzy hybrid with biological knowledge to infer strong regulatory relationships (activation, inhibition, or no effect) in the form of fuzzy rules. They proposed six-layered, two-input and one-output neuro-fuzzy network (Fig. 7). The first layer represents an input linguistic variable, where values are transferred to another layer without any computation. In the $2^{nd}$ and $3^{rd}$ layer, input linguistic variable is converted into fuzzy output using different membership functions. In the $4^{th}$ and $5^{th}$ layer, fuzzy rules are inferred, and finally $6^{th}$ layer is used for the defuzzification.

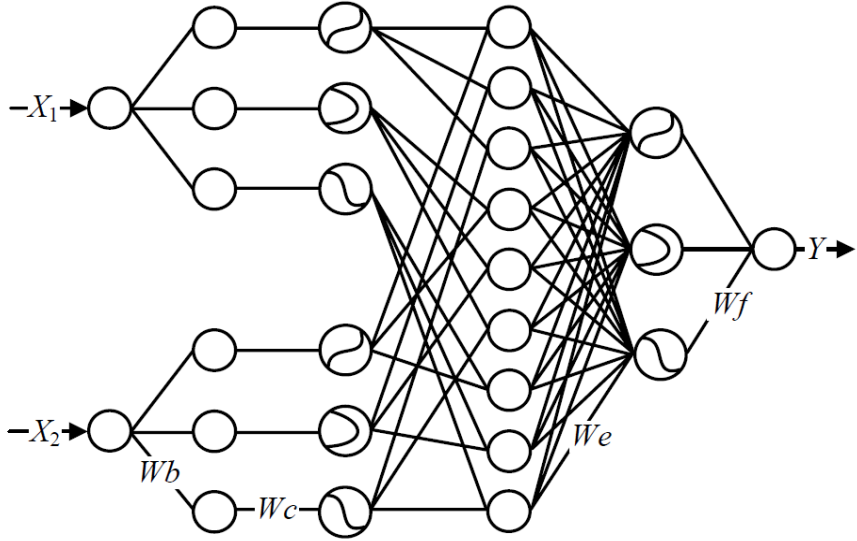

**Fig. 7** Six-layered Neuro-Fuzzy GRN inference model proposed by Liu et al. (2011)

Manshaei et al. (2012) proposed Hybrid Rule-Based Neuro-fuzzy (HRBNF) algorithm for GRN reconstruction from medium or small number of available measurements. The HRBNF algorithm follows multi-stages of decision making to infer gene-gene interactions using rules which govern the gene expressions. The algorithm consists of five stages, starting with gene expression training (stage 1) to extract set of rules (stage 2), sort it (stage 3), and compare the rules for GRN analysis (stage 4), and finally modeling final GRN (stage 5).

Vineetha et.al. (2012) proposed TSK-type six-layered recurrent neural-fuzzy method to infer regulatory relationship between genes and reconstructed GRN for circulating plasma network using colon cancer gene expression data. Here, two phases of learning, namely structural and parameter learning, is applied. The structural learning was used for input-output space partition, fuzzy if-then rules, and feedback structural identification. The parameter learning was applied for tuning network's free parameters. Neural network with weighted fuzzy membership function (NEWFM) combines inference and learning capabilities into a neuro-fuzzy system. NEWFM approach divides layers using bounded sum of weighted fuzzy membership function which are learned from the network. Yoon et al. (2015) applied NEWFM to model GRN which untangles model complexity and simplify fuzzy inference. It also improves the reconstruction accuracy without compromising dynamic regulatory cycle. Wang et al. (2016) also applied NEWFM to model the relationships between genes in yeast cell cycle. Their approach consists of four stages: (i) Learning using NEWFM, (ii) regulator selection, (iii) activator/repressors classification, and (iv) GRN reconstruction. Through gene preprocessing, two kinds of features were selected for learning neural network. For regulator selection, genes having the best or worst effects on the target genes were considered. Their method performed better than Time-delay ARACNE method (Zoppoli et al., 2010) for yeast cell cycle data, but discovery rate of the repressors were found to be low. Review of neuro-fuzzy model of GRN and meta-heuristic algorithms used to learn structure and parameters of GRN can be found in Biswas & Acharyya (2016).

## 4.5 Fuzzy Evolutionary

Evolutionary algorithms (EAs) are intelligent optimization techniques inspired by Darwinian theory of evolution "survival for the fittest". These techniques mimic the natural evolution of the species in order to develop new search methods which are robust, noise tolerant and search for solutions in an almost infinite search space (Linden & Bhaya, 2007; Raza & Parveen, 2013a). In EAs, searching for a solution within a population is carried out from a single point and a competitive selection is done. The solutions with high fitness values are recombined with other solutions, and then mutated to generate new solutions space. Some of the popularly known EAs are genetic algorithms (GA), gene programming, and evolutionary programming (EP). The GA focuses on optimization of combinatorial problems, while GP is used for evolving computer programs and EP for optimizing continuous functions without using recombination (Raza & Parveen, 2013a; Raza, 2016a). EAs are hybridized with other intelligent techniques such as ANNs for GRN inference problems, which is beyond the scope of this paper. However, here we focus on the fuzzy logic hybridized with EAs for GRN inference and modeling.

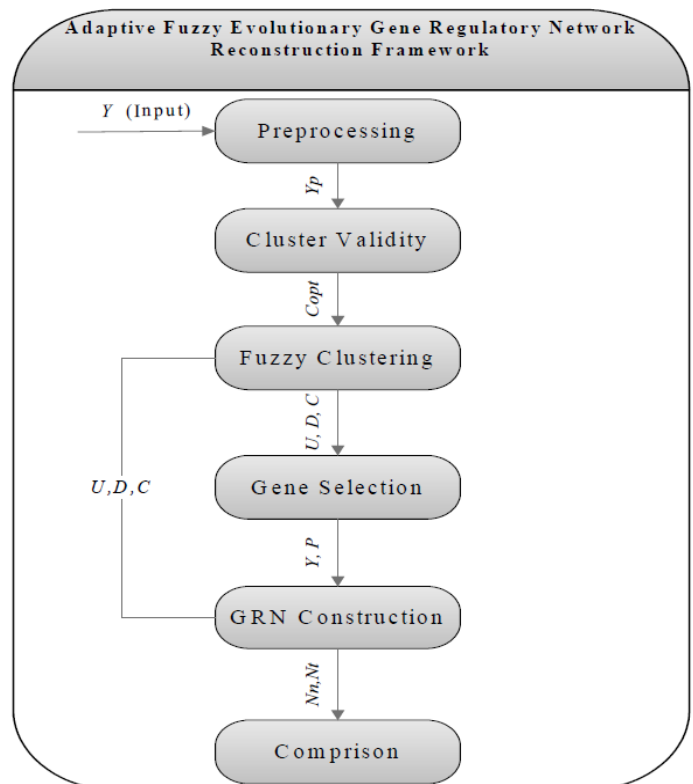

**Fig. 8** Adaptive Fuzzy Evolutionary GRN reconstruction framework proposed by Sehgal et al. (2006)

Fuzzy logic is easily combined with the evolutionary computing to optimize some of its parameters. One of the hybrid of fuzzy with evolutionary computing, called Adaptive Fuzzy Evolutionary GRN Reconstruction (AFEGRN), is applied for modeling GRN by Sehgal et al. (2006). The AFEGRN framework has six steps as depicted in Fig. 8. The preprocessing step removes noise and outliers in the data, which is followed by number of clusters estimation step. After cluster estimation, data is clustered using fuzzy c-means (FCM) clustering, followed by significant gene selection using Between Group to within Group Sum of Squares (BSS/WSS) method. Finally, GRN is constructed using simple statistical technique such as spearman ranked correlation. This framework is suitable for the reconstruction of GRN for two different samples such as control and disease. Linden & Bhaya (2007) used GP and fuzzy logic hybrid to extract gene regulatory rules from gene expression profile and the method also facilitate incorporating prior biological knowledge into the model. To deal with the "*curse of dimensionality of problem*", Linden & Bhaya (2007) first group the co-regulated genes based on their co-expression profiles, and then GP is applied to evolve the network. They used reverse polish notation to represent the rules within the chromosome structure, and only three operators, such as AND, OR and NOT, were used which led to smaller, simpler and easily understandable results.

*4.6 Fuzzy Petri Net*

A Petri net is a mathematical modeling technique which describes discrete event in dynamic systems. It is a directed bipartite multi-graph consisting of two types of nodes, called *places* $P=\{p_1, p_2,..., p_n\}$ (depicted as circles) and *transitions* $T=\{t_1, t_2, ..., t_m\}$ (depicted as rectangle), and directed edges which connects only nodes of different types and weighted by natural numbers (Sackmann et al., 2006). Places (P) stand for passive system elements (e.g. states, conditions, or biological macromolecules such as genes, proteins or metabolites) and transitions (T) represent active systems elements (e.g. events, chemical reactions, etc.). The directed edges (*arcs*) do not only describe the causal relation between active and passive elements but also define the effect of a reaction that specify the quantity of substrate consumed and quantity of product produced during the reaction (called *firing* of a transition). The entity state is defined by the *tokens* which represent the *marking* of the place. A general architecture of FPN is shown in Fig. 9.

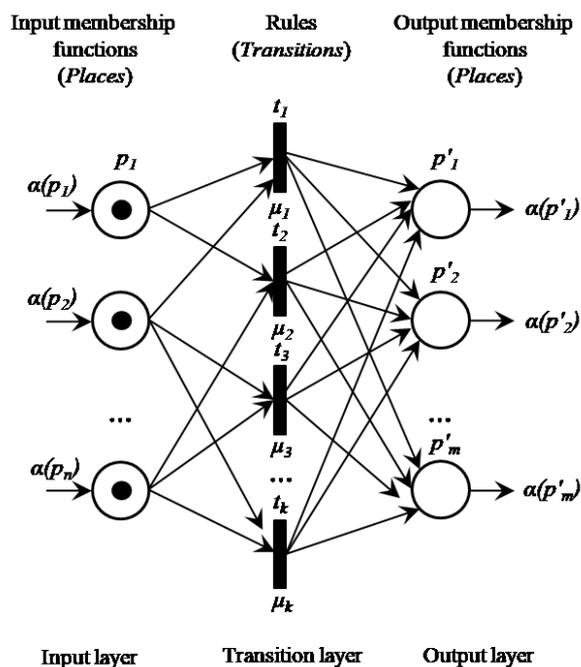

**Fig. 9** A general three layer architecture of FPN consisting of *n* input places at input layer, *k* hidden transitions at transition layer, and *m* output places at output layer, the token function $\alpha$ associated with places, and certainty factor $\mu_i$ associated with transitions.

Fuzzy Petri net (FPN), expanded from Petri net, has a token function ($\alpha_i \in [0,1]\ \forall\ i = \{1,2,...m\}$) associated with places, and a certainty factor (CF) ($\mu_j \in [0,1]\ \forall\ i = \{1,2,...m\}$) associated with transition. FPN is a promising modeling technique for large and complex systems which has capabilities for fuzzy knowledge representation and reasoning. The FPNs have been applied for modeling and simulation of GRN (Hamed et al., 2010a; Hamed et al., 2010b; Küffner et al., 2010; Hamed, 2013; Hamed, 2017; Li et al., 2017).

Hamed et al. (2010a) proposed FPNs based GRN model for searching activator/repressor regulatory relationship under gene triplets framework in gene expression. They presented FPNs based model of GRN as 9-tuple (Fig. 10):

$$FPN = \{P, T, D, I, O, f, \alpha, \beta, \lambda\} \qquad (12)$$

where,

$P=\{p_1, p_2, ..., p_n\}$ finite set of places,

$T=\{t_1, t_2, ..., t_n\}$ finite set of transitions,

$D=\{d_1, d_2, ..., d_n\}$ finite set of proposition,

$I$: input incidence matrix,

$O$: output incidence matrix,

$f=\{\mu_1, \mu_2, ..., \mu_m\}$, where $\mu_i \in [0,1]$ is the certainty factor of the reliability of rule $R_i$.

$\alpha: P \rightarrow [0, 1]$ is a function that assigns a token value between [0, 1],

$\beta: P \rightarrow D$ is an association function performing bijective mapping from places to propositions,

$\lambda: T \rightarrow [0, 1]$ is the function which assigns a threshold $\lambda_i$ to transition $t_i$.

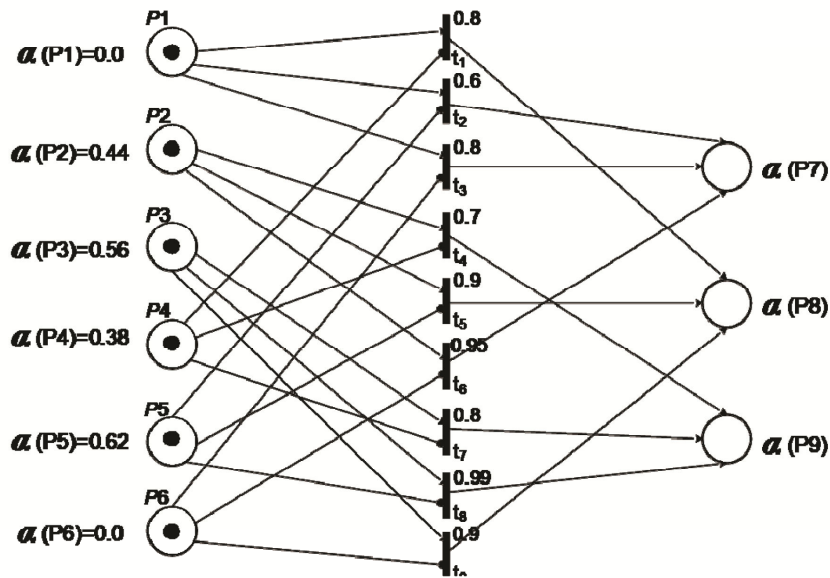

**Fig. 10** FPN model of GRN proposed by Hamed et al. (2010a). The example consists of nine places $P=\{p_1, p_2, ..., p_9\}$, transitions $T=\{t_1, t_2, ..., t_9\}$, the initial degree $\alpha=\{0, 0.44, 0.56, 0.38, 0.62, 0, 0, 0, 0\}^T$, initial marking vector $M_0=\{1, 1, 1, 1, 1, 1, 0, 0, 0\}^T$, and certainty factor $\mu_i = \{0.8, 0.6, 0.8, 0.7, 0.9, 0.95, 0.8, 0.99, 0.9\}$.

Hamed et al. (2010b) extended their previous FPN model by incorporating the concept of hidden fuzzy transition (HFT) as a new type of transition which allows wider search spaces to infer regulatory relationship. They define FPN model as 10-tuple (Fig. 11),

$$FPN = \{P, T, D, I, O, f, \alpha, \beta, \lambda, HFT\} \quad (13)$$

where, $HFT = \{hft_1, hft_2\}$. The model assumed the following fuzzy production rules based on initial degree α (Hamed et al., 2010b),

$R$1: If $d_1$ and $d_5$ then $d_7$ (CF = 0.8)
$R$2: If $d_1$ and $d_6$ then $d_7$ (CF = 0.6)
$R$3: If $d_2$ and $d_6$ then $d_7$ (CF = 0.8)
$R$4: If $d_1$ and $d_4$ then $d_8$ (CF = 0.7)
$R$5: If $d_2$ and $d_5$ then $d_8$ (CF = 0.9)
$R$6: If $d_3$ and $d_6$ then $d_8$ (CF = 0.95)
$R$7: If $d_2$ and $d_4$ then $d_9$ (CF = 0.8)
$R$8: If $d_3$ and $d_4$ then $d_9$ (CF = 0.99)
$R$9: If $d_3$ and $d_5$ then $d_9$ (CF = 0.9).

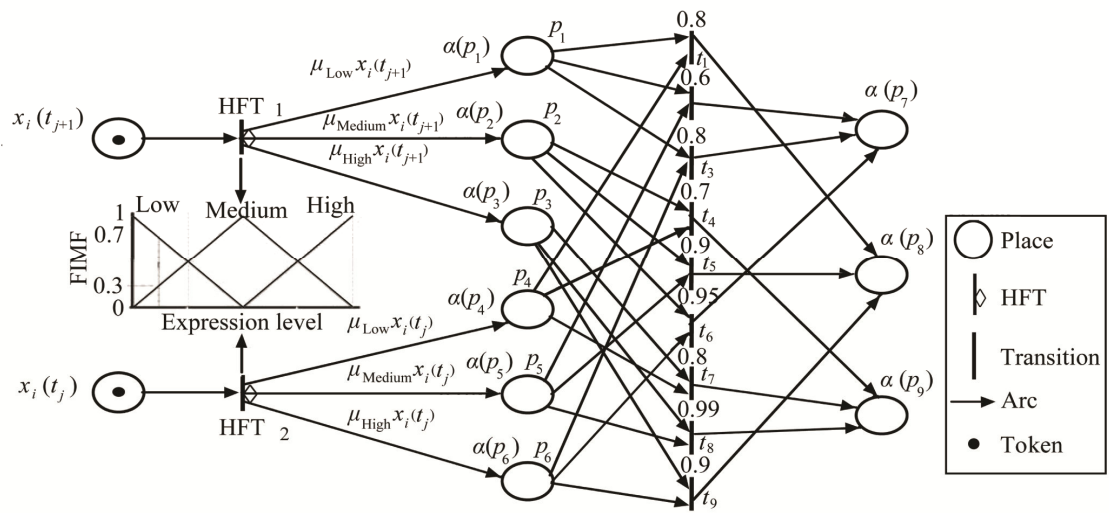

**Fig. 11** FPN model of GRN with HFT transition proposed by Hamed et al. (2010b). The example consists of nine places $P=\{p_1, p_2, ..., p_9\}$, transitions $T=\{t_1, t_2, ..., t_9\}$, the initial degree $\alpha=\{0, 0.48, 0.52, 0.4, 0.6, 0, 0, 0, 0\}^T$, initial marking vector $M_0=\{1, 1, 1, 1, 1, 1, 0, 0, 0\}^T$, and certainty factor $\mu_i = \{0.8, 0.6, 0.8, 0.7, 0.9, 0.95, 0.8, 0.99, 0.9\}$.

Another FPN based model was proposed by Küffner et al. (2010) which was ranked as the best performer method for DREAM4 competition of size ten networks. The model of Kuffner et al. (2010) utilizes diverse datasets, such as knockout and knockdown mutations, multifactorial, and time course data, into the model.

Since fuzzy production rules (FPRs) are applied for knowledge representations in FPN, it becomes essential to use composite conjunctive FPRs whose antecedent consists of more than one proposition. To deal with the relative degree of importance of a proposition contributing to its consequence, a "local weight" vector may be used. This idea of "local weight" in FPRs (also called weighted FPRs) was applied by Hamed (2013) to successfully develop FPNs based GRN model. The weighted FPRs reduce the undesirable effects on subsequent part of FPN, and helps interpreting the linguistic meaning in better way.

Although Petri net and FPN have capability to model concurrent processes but it suffers from hierarchical structuring which limits its application on large-scale networks. Also, it has no types, no modules, and allows only one kind of tokens. Color Petri net (CPN) is a high-level Petri net with a hierarchical structure where it is possible to incorporate data types, complex data manipulation, and each token has data value attached to it, called the token color (Jensen, 2013). FPN can be combined with CPN, called fuzzy color Petri net (FCPN), to take advantages of both. Li et al. (2017) applied FCPN to integrate reverse reasoning into the GRN model which simplifies the network size, and influence degree of specific genes on the target gene is predicted, and causal relationship between genes is simulated effectively.

FPN is a powerful modeling tool for FPRs based knowledge systems, but it lacks the learning capability. The lack of learning capability makes the FPN unsuitable for modeling uncertain knowledge systems. Fuzzy neural Petri net (FNPN), a neural extension to FPN, has fuzzy neuron components as a subnet model where parameters of FPRs can be learnt and trained (Xu et al., 2007). Hence, FNPN is suitable for modeling uncertain knowledge systems. Hamed (2017) proposed FNPN based model of GRN which can deduce the dynamic information with self-learning capacity.

Although, FPN is graphical tool which allows structuring a rule-based fuzzy reasoning system, however it needs both confidence degree of rule and truth degree of preposition a priori which needs the experience of the experts.

## *4.7 Fuzzy Answer Set Programming*

Answer Set Programming (ASP) is a declarative programming paradigm based on answer set (stable model) semantics of logic programming which is designed for difficult search problems, primarily NP-hard problems and useful in knowledge-intensive applications (Lifschitz, 2008; Eiter et al., 2009). Since ASP is oriented towards solving NP-hard search problems, therefore it is reduced to computing stable models and answer set solvers. Fuzzy logic can be hybridized with ASP into single framework, called fuzzy answer set programming (FASP). FASP offers features of both the fields: from SAP, it takes truly declarative reasoning capabilities, while fuzzy logic gives flexibility of interpretation of beyond sharp principles of classical logic (Van Nieuwenborgh et al., 2007).

The first application of FASP to model the dynamics of GRN and to find attractors of its nodes was carried out by Mushthofa et al. (2016). They applied FASP to model the dynamics of multi-valued GRNs (extension of Boolean network) and computed the multi-valued activation of each node. Their work demonstrated that multi-valued networks in any *k* can be successfully encoded using FASP which can reasonably capture underlying assumptions needed in the modeling of GRNs. Further, authors extended FASP applications to randomly generated artificial data as well, and developed a tool, called FASPG (Mushthofa et al., 2018). The workflow of FASPG program is shown in Fig. 12. The network specification in terms of regulatory relationships consists of all the possible combination of every node regulating it, which is represented using fuzzy logic formulas under Lukasiewicz semantic. Finding suitable formulas which fits certain input-output relationship specification is not straightforward. FASPG invokes FASP program to compute attractors of GRN, which further invokes ASP solver. The input to FASPG program is the description of a GRN comprising of, (i) number of genes ($n$), (ii) number of activation levels of each gene ($k$), and (iii) an input-output specification – which is a set of assignments for a gene given all possible combinations of genes regulating it. For instance, if a gene $g$ is regulated by $n$ genes ($e_1, e_2, ..., e_n$), the input-output specification for $g$ would be a table of $k^n$ rows, each comprising of a possible combination of values of $e_i$ and a corresponding value of $g$ (Mushthofa et al., 2018). Both the programs FASPG (http://github.com/mushthofa/faspg) and FFASP (http://github.com/mushthofa/ffasp) are available on github. The application of FASP can be further explored on some real biological networks to find the best possible attractors.

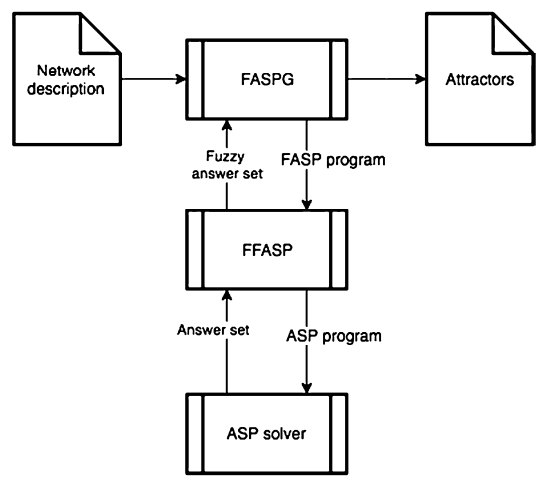

**Fig. 12** The workflow of FASPG program to find possible attractors (Mushthofa et al., 2018)

### *4.8 Other Fuzzy Hybrid*

Fuzzy logic has also been hybridized with several other computational intelligence methods such as union rule configuration (Sokhansanj & Fitch, 2001), exhaustive search (Sokhansanj et al., 2004), network component analysis (Bakouie & Moradi, 2007), Bayesian networks (Wang et al., 2008), and so on.

Sokhansanj & Fitch (2001) applied fuzzy logic with Union Rule Configuration (URC) for modeling and simulation of gene regulation. The URC is applied to avoid combinatorial explosion problem in the fuzzy rules, and therefore it can be used to model complex biological system. Sokhansanj et al. (2004) applied linear fuzzy GRN model and extract gene regulation by exhaustive search. The model considered potential interactions between 12 genes from yeast cell cycle, and hence rule for a target gene may have any combination and number of 11 other genes. Since input gene influences other gene by any one of the 27 possible fuzzy rules, there would be ~1016 possible number of rules for each of the 12 genes. This model recovers both direct and indirect interactions by best-fitting GRN models and exhaustive rule search.

For computing low-dimensional data using Principal Component Analysis (PCA) and Independent Component Analysis (ICA), the underlying network structures are ignored and data decomposition is solely based on a priori constraint. Network Component Analysis (NCA) is an emerging method for inferring hidden regulatory interactions. NCA tries to decompose a given matrix E into two matrix $E = AR$, where unlike PCA and ICA, some of the entries of R are constrained to 0, where R is a regulatory layer. By taking the advantage of partial network connectivity knowledge, fuzzy based NCA has been applied to reconstruct GRN by Bakouie & Moradi (2007). NCA organizes multi-dimensional gene expression data [E] into N genes and M samples to reconstruct a model such as,

$$[E] = [A] \times [R] \tag{14}$$

where, matrix $[A]_{N \times L}$ encodes connectivity strength between regulatory layer and output signal, matrix $[R]_{L \times M}$ consists of L samples regulatory signals. Since L<N, hence it reduces the dimensionality. Fuzzy clustering algorithm, such as fuzzy c-means, has also been found several applications in GRN reconstruction process and mostly used as data preprocessing step. They hybridized fuzzy clustering with NCA to infer regulatory interactions. Bayesian Networks (BNs) are directed acyclic graph (DAG) which can also be used to represent gene regulatory relationships. BN is suitable for small networks however, if number of genes are large, learning a good BN is very difficult due to exponential growth of search space. Wang et al. (2008) combined fuzzy clustering with Bayesian Networks to predict GRN. The BN is used to predict GRN and fuzzy clustering algorithm is applied to reduce the search space. Also, BN learning suffers from dimensionality and over-fitting problem due to the layout of microarray data. To deal with this problem in GRN modeling, Njah & Jamoussi (2015) applied a fuzzy ensemble clustering approach which outputs small and highly inter-correlated partitions of genes. After estimation of optimal number of clusters, an ensemble method is applied to construct a consensual partition of the training dataset. For learning BNs of each partition (sub-BNs), Njah & Jamoussi (2015) applied a weighted committee based structure algorithm. Further, sub-BNs are assembled through common genes.

Barman et al. (2016) developed adaptive ANN and self-organising map (SOM) based GRNs of Hepatitis C virus infection effect on Huh7 hepatoma cell time-series gene expression data, where they applied fuzzy C-means clustering for the identification of cluster centres before reconstructing GRNs.

**Table 1** Fuzzy logic and its hybridized methods for GRN inference

**List of Abbreviations:** Ant Colony Optimization (ACO), Artificial Neural Network (ANN), Bayesian Network (BN), Compress Sensing (CS), Color Petri net (CPN), Dynamic Multi-Agent GA (dMAGA), Evolutionary Algorithm (EA), FCM with Decomposition (FCMD), Fuzzy Production Rule (FPR), Genetic Algorithm (GA), Genetic Programming (GP), Hidden Fuzzy Transition (HFT), Least Absolute Shrinkage and Selection Operator (LASSO), Mematic Algorithm (MA), Mutual Information (MI), Network Component Analysis (NCA), Weighted Fuzzy Membership (WFM), Ordinary Differential Equation (ODE), Union Rule Configuration (URC).

| S.No. | Fuzzy logic models | Descriptions | Datasets | References |
|---|---|---|---|---|
| *Classical Fuzzy Logic (FL)* | | | | |
| 1. | FL | Finds gene triplets (Activators-Repressors-Targets) | Yeast | Woolf & Wang (2000) |
| | | A quantitative fuzzy logic model which cope with unknown kinetic data. | Three-gene repressilator | Bordon et al. (2005) |
| 2. | FL + Clustering | To reduce number of possible triplets, clustering is used as preprocessing steps. Also introduced co-activators and co-repressors in the GRN model. | Yeast | Ressom et al. (2003a); Ressom et al. (2003b) |
| | | Clustering as preprocessing step to group genes based on their gene expression similarity. | Yeast | Ram et al. (2006) |
| | | Fuzzy network module extraction (FUMET) having highly co-expressed genes | Yeast, Human, Rat CNS | Mahanta et al. (2014) |
| *Fuzzy Cognitive Map (FCM)* | | | | |
| 3. | FCM | Models the causal relationship between biomolecules. | Arabidopsis | Du et al. (2005) |
| 4. | FCM+ACO | FCM is used to represent GRN and ACO to learn FCM. It supports problem decomposed to make the algorithm scalable. | DREAM4 | Chen et al. (2012) |
| 5. | FCM + decomposed GA | GA is applied to learn large-scale FCMs | DREAM3, DREAM4 | Chen et al. (2015) |
| 6. | FCM + MA + ANN | MA is used to find regulatory interactions and ANN to estimate GRN interaction strength | DREAM3, DREAM4 | Chi & Liu (2016) |
| 7. | FCM + LASSO | Decompose FCM learning problem into sparse signal reconstruction problem and robustly train FCM from noisy data | DREAM3, DREAM4 | Wu & Liu (2016) |
| 8. | FCMD + dMAGA | Train large-scale FCMs | DREAM3, DREAM4 | Liu et al. (2016) Liu et al. (2017) |
| 9. | FCM + CS (CS-FCM) | Decompose FCM learning problem and performs well for network having more than 1,000 nodes | DREAM3, DREAM4 | Wu & Liu (2017) |
| 10. | FCM + MI + MA | Train large-scale FCMs | DREAM3, DREAM4, Real-life data from literature | Zou & Liu (2017) |
| *Dynamic Fuzzy Models (DFM)* | | | | |
| 11. | DFM | Incorporate structural knowledge into the model and infers gene interactions in the form of fuzzy rules. | SOS DNA repair network with structural knowledge | Sun et al. (2010) |
| *Neuro-Fuzzy (NF) and Neuro-Fuzzy Inference System (NFIS)* | | | | |
| 12. | NFIS | Model consists of two modules: neuro-fuzzy inference module and evolutionary strategy learning | Simulated data | Jung & Cho (2007) |
| 13. | Recurrent NF + EA | Evolutionary algorithm is used to train the network. It is able to find complex causal relationship between genes. | S. Cerevisiae, E. Coli | Maraziotis et al. (2007) |
| 14. | NFIS+ ODE | Combines features from ODE based FIS model, trained with ANN | Lac Operon in E. Coli | Munoz et al. (2009) |
| 15. | Recurrent NF | Determined the gene-gene interactions parameters as weights between neurons of neural network. | SOS DNA repair network | Datta et al. (2009) |
| 16. | NF | Incorporates biological knowledge into the model to infer strong regulators as fuzzy rules | Cell cycle of yeast | Liu et al. (2011) |
| 17. | Recurrent NF | TSK-type six-layered recurrent neuro-fuzzy model | Circulating plasma network from colon cancer | Vineetha et al. (2012) |
| 18. | NF + WFM (NEWFM) | Combines learning and inference capability into a neuro-fuzzy system. | Cell cycle of yeast | Yoon et al. (2015) |
| 19. | NEWFM | Model consists of four stages: learning using NEWFM, regulator selection, activator/repressors classification, and GRN reconstruction. | Cell cycle of yeast | Wang et al. (2016) |
| *Fuzzy Evolutionary Hybrid* | | | | |
| 20. | FL + GP | Extract GRN and capable of incorporating prior biological knowledge. | Arabidopsis thaliana cold response; Rat central nervous system | Linden & Bhaya (2007) |
| 21. | FL + EA + Clustering | A six-step framework suitable for GRN reconstruction for two samples such as control and disease | Breast cancer and normal dataset | Sehgal et al. (2006) |
| 22. | ODE + FIS (ODE-FIS) | Combine features of ODE to fuzzy inference system based GRN model. | Lac Operon in E. Coli | Munoz et al. (2009) |
| 23. | Hybrid Rule-Based Neuro-Fuzzy (HRBNF) | Follow four-stage decision making to infer regulatory interactions from medium or small number of samples. | Cell cycle of yeast | Manshaei et al. (2012) |
| *Fuzzy Petri Net (FPN)* | | | | |
| 24. | FPN | Extract activator/repressor regulatory relation under gene triplet framework | Simulated data | Hamed et al. (2010a) |
| 25. | FPN + HFT | Introduce HRF as new transition type, allowing wider search space | Simulated data | Hamed et al. (2010b) |
| 26. | FPN | Incorporates diverse dataset such as knock-out and know-down mutation, multi-factorial and time-series. Best performer in DREAM4 challenge of size 10 network | DREAM4 | Kuffner et al. (2010) |
| 27. | FPN + FPR | Applied FPR to introduce local weight which helps interpreting linguistic meaning in better way | Simulated data | Hamed (2013) |
| 28. | FPN + CPN (FCPN) | Integrate reverse reasoning into FPN using CPN. Also, it has structural hierarchical, data types and each token has a value attached to it. | DNA sequence of 6 bases | Li et al. (2017) |
| 29. | FPN + ANN (FNPN) | A neural network extension to FPN which allow parameters learning | Simulated data | Hamed (2017) |
| *Fuzzy Answer Set Programming (FASP)* | | | | |
| 30. | FASP | Models the dynamics of multi-valued GRN | Randomly generated artificial data | Mushthofa et al. (2016); Mushthofa et al., 2018 |
| *Other Fuzzy Hybrids* | | | | |
| 31. | FL + URC | URC is applied to avoid combinatorial explosion problem in fuzzy rules | Lac Operon in E. Coli | Sokhansanj & Fitch (2001) |
| 32. | FL + Exhaustive Search | Extract gene regulation by exhaustive search | Network of 12 genes from cell cycle | Sokhansanj et al. (2004) |
| 33. | FL + NCA | NCA provides problem decomposition ability and extracts hidden regulatory interactions | Ternary expression data | Bakouie & Moradi (2007) |
| 34. | Fuzzy clustering + BN | Fuzzy clustering reduces search space and BN is used to predict GRN | Cell cycle of yeast | Wang et al. (2008) |
| 35. | Fuzzy ensemble clustering + BN | Finds optimal number of clusters and construct a partition of training dataset using ensemble method. BN is used for training a GRN. | Iris, Ionosphere, Glass, Wine | Njah & Jamoussi (2015) |

## 5. Discussion and Conclusion

Gene regulatory network inference (GRNI) from high-throughput gene expression data is a well-posed challenge from last few decades. Several computational methods have been proposed ranging from simple statistical to sophisticated computational intelligence approaches. Among these approaches, fuzzy logic theory has lots of potential applications in different areas of bioinformatics, including GRNI. Fuzzy logic is capable to represent nonlinear systems and incorporate domain knowledge using fuzzy rules. Some of the advantages of fuzzy logic in gene expression studies are (i) it extracts trends rather than precise values, and thus it can inherently handle noises in the data; (ii) its predicted results are easily interpretable, and (iii) it is computationally efficient and scalable. An initial successful attempt to apply fuzzy logic for GRNI was done by Woolf and Wang (2000) where gene triplets consisting of activators, repressors, and targets were identified. Later on, it was improved by several researchers in terms of accuracy and reduction in computational cost.

This paper presented fuzzy logic and its hybridization with other computational techniques for GRNI such as fuzzy cognitive maps (FCMs), dynamic fuzzy modeling, neuro-fuzzy, neuro-evolutionary, fuzzy Petri nets, fuzzy answer set programming and other fuzzy hybrids. FCMs combine features from fuzzy logic and ANN and works well for modeling complex processes including GRN. It has the capability to cope with lack of quantitative information as how various biomolecules interact. The genes are represented as causal fuzzy sets and degree to which they are dependent on each other. Traditionally, construction of FCMs relies on domain knowledge but it can also be learn automatically and discover the domain knowledge. However, approaches to learn FCMs automatically is limited to small-scale networks, very low accuracy, high computational complexity, and very high density of learned FCMs. Few attempts have been made to learn large-scale FCMs using various decomposition methods including decomposed genetic algorithm, ACO, differential evolution, PSO, LASSO, Compressed Sensing, multi-agent genetic algorithm (dMAG), mutual information based two-phase memetic algorithm (MIMA), and memetic algorithm combined with ANN (MANN).

Dynamic fuzzy modeling approach has the capability to incorporate the prior structural knowledge to the GRN model and infer gene interactions as fuzzy rules. Neuro-fuzzy is one of the most widely applied hybrid approach for GRNI which combines the learning and adaptation feature of ANN and knowledge representation through fuzzy logic. Several fuzzy-evolutionary hybrid approaches have been proposed for GRNI and network parameter optimization. Most of these approaches are multi-layer models, starting first layer as data preprocessing and clustering to final layer as predicted GRN with optimized regulatory interactions parameters. Fuzzy Petri net (FPN) is also a promising modeling technique for large and complex systems which has capabilities for fuzzy knowledge representation and reasoning. Several FPNs based approaches developed for modeling and simulation of GRN. Although FPNs have capability to model concurrent processes but it suffers from hierarchical structuring which limits its application on large-scale networks. Also, it has no types, no modules, and allows only one kind of tokens. Hence, fuzzy color Petri net (FCPN) is explored for its applications in GRNI which supports hierarchical structure, data types, complex data manipulation, and each token has data value attached to it. The lack of learning capability makes the FPN unsuitable for modeling uncertain knowledge systems. Therefore, fuzzy neural Petri net (FNPN) was applied to learn parameters of FPRs. Although, FPN is graphical tool which allows structuring a rule-based fuzzy reasoning system, however it needs both confidence degree of rule and truth degree of preposition a priori which needs the experience of the experts. Fuzzy answer set programming (FASP) is a declarative programming paradigm which is suitable for difficult search problems and useful in knowledge-intensive applications. FASP was also explored to model the dynamics multi-valued of GRN and computed multi-valued activation of each gene.

Fuzzy logic based other hybridized computational methods was also proposed for GRNI including union rule configuration (URC), exhaustive search, network component analysis (NCA), Bayesian networks (BN), and so on. The URC were applied to avoid combinatorial explosion problem in the fuzzy rules, while exhaustive search recovers both direct and indirect regulations. Similarly, for computing low-dimensional data, NCA tries to decompose data and takes the advantage of partial network connectivity knowledge. BN is suitable for small networks, however, if number of genes are large, learning a good BN is very difficult due to exponential growth of search space. Therefore, to reduce the search space in BN, fuzzy clustering can be applied as preprocessing step. Also, BN learning suffers from dimensionality and over-fitting problem due to the layout of microarray data. To deal with this problem in GRN modeling, fuzzy ensemble clustering approach is used which outputs small and highly inter-correlated partitions of genes. One of the serious drawbacks of most of the GRNI methods is that these have been tested on simulated gene expression data available from DREAM challenge and lacks it rigorous testing on real gene expression data. Also, these methods have been either tested on small- or mid-size networks. To deal with large-scale networks we need better and robust decomposition approaches. Further, only gene expression data would not be enough to infer GRN accurately. Therefore, we need better data integration techniques so that multi-omics data (such as miRNA expression, ChIP-seq/ChIP-ChIP, mutation data such as SNPs, CNVs, GO annotations, protein-protein interaction and gene-disease association data) can be easily integrated into the GRN inference model to achieve accurate and reliable results.

**Conflict of Interest**

The author declares that there is no any conflict of interest.